\documentclass[12pt]{article}
\usepackage{amsmath}
\usepackage{amssymb}
\usepackage{graphicx}
\newcommand{\be}{\begin{equation}}
\newcommand{\ee}{\end{equation}}
\newcommand{\ba}{\begin{eqnarray}}
\newcommand{\ea}{\end{eqnarray}}
\newcommand{\ds}{\displaystyle}

\begin{document}
\vspace{1cm}

\begin{center}
\bf Quasi-Two Dimensional Diluted Magnetic Semiconductors \\with Arbitrary Carrier Degeneracy\\
\end{center}
\bigskip

\centerline{E.Z. Meilikhov, and R.M. Farzetdinova}
\medskip
\centerline{\small\it Russian research center “Kurchatov Institute”,
123182 Moscow, Russia}
  \vspace{1cm}

  \centerline{
\begin{tabular}{p{15cm}}
\footnotesize \hspace{15pt}In the framework of the generalized mean
field theory, conditions for arising the ferromagnetic state in a
\emph{two-dimensional} diluted magnetic semiconductor and the
features of that state are defined. RKKY-interaction of magnetic
impurities is supposed. The spatial disorder of their arrangement
and  temperature alteration of the carrier degeneracy are taken into
account.
\end{tabular}
} \vspace{1cm}

Diluted magnetic semiconductors, such as
Ga$_{1-x}$Mn$_x$As~\cite{1}, are  broadly investigated in connection
with their potential for new electronics developments and,
especially, spintronics.  For explanation of the ferromagnetism in
those compounds the known RKKY-mechanism of the indirect exchange
interaction is recruited ~\cite{1,2} which leads to the correct
estimating the Curie temperature in the framework of the traditional
mean field theory.  Mn-atoms (with concentration $N_{\mu}$)
substituting for Ga-atoms introduce in the system the own magnetic
moments and, in addition, as acceptors deliver free holes (with
concentration $n$). It is precisely those holes become to be
carriers responsible for the interaction. However, the equality  of
the concentrations $n=N_{\mu}$ keeps only at low Mn-concentrations
($x\lesssim0.05$)~\cite{1,2}, so that the carrier concentration is
usually less than the concentration of magnetic impurities:
$n=\gamma N_\mu$ where the coefficient of the impurity “efficiency”
$\gamma<1$ and falls\footnote{One could control the relative hole
concentration (i.e., the  $\gamma$-value) by simultaneous
introducing non-magnetic acceptors (for instance, Be~\cite{3,4}) or
choosing the temperature of the film growth~\cite{5}.} with
increasing $N_\mu$.

Nevertheless, the concentration of magnetic impurities, delivering
carriers, in actual systems is usually so high that the impurity
band is formed which at $x\gtrsim0.01$ merges into the valence
band~\cite{5}. Even though, the carrier concentration occurs to be
not so high that one could consider them as highly degenerated ones
within the whole (being of interest) range of relatively high
temperatures. Furthermore, it is important that in that range the
carrier concentration is almost independent of the temperature:
$n=\gamma N_\mu\approx\mbox{Const}$.

Although two-dimensional structures  represent the most natural
systems for the embedding in the traditional semiconductor
technology, almost all theoretical works are dealt with the
three-dimensional systems of degenerate carriers. Similarly,  most
of experimentally studied systems are three-dimensional ones.

The objective of the present paper is considering magnetic features
of \emph{two-dimensional} semiconductor systems with magnetic
impurities  interacting  by RKKY-mechanism via carriers of
\emph{arbitrary degeneracy}. That problem has been recently
considered in the paper~\cite{6} where it has been shown that
reducing the system dimension (from 3D to 2D) results in the
significant lowering of the Curie temperature (under equivalent
parameters). We think though that there are some inexactitudes in
the paper. Firstly, authors~\cite{6} has neglected of the
temperature dependence $\varepsilon_F(T)$ of the carrier Fermi
energy that is inadmissible under the intermediate degeneracy when
$\varepsilon_F/k_B T\sim1$ (and, all the more, at
$\varepsilon_F<0$). In addition, the disorder in the arrangement of
magnetic impurities has been considered in the framework of the mean
field theory (continual or the so called lattice one) where the
averaged impurity configuration is \emph{ordered} one and the
\emph{scattering} of local effective magnetic fields is not
considered. As we demonstrate  below, taking into account the
mentioned things influences the results significantly.

The general expression for RKKY interaction in two-dimensional
systems is known~\cite{7,8,9}, however the relevant results refer to
the degenerate carriers and have to be generalized for the case of
the intermediate degeneracy $|\varepsilon_F/k_BT|\sim 1$
($\varepsilon_F$ is the carrier Fermi energy).

In a few papers~\cite{10, 11} (and later in~\cite{12}) the systems
with non-degenerate carriers have been considered and it has been
shown that the energy of RKKY interaction in the non-degenerate case
is exponentially falls with the distance (in contrast to the
degenerate case where the fall is  power behaved):
$J\propto\exp(-r^2/\lambda_T^2)$, where
$\lambda_T=\hbar/(2mk_BT)^{1/2}$ is the mean (thermal) de Broglie
carrier wave length. However, those papers, first, refer to the
three-dimensional systems and, second, they lack of taking into
account the random arrangement of magnetic impurities.

It is known that the traditional mean field theory does not provide
the adequate description of a \emph{disordered} (random) system of
magnetic moments~\cite{13}. In the present paper, we shall use the
generalized mean field theory~\cite{14} for systems with the
indirect interaction of magnetic impurities taking into account the
randomness of their spatial arrangement. We shall use Ising
approximation and suppose that the indirect coupling between
magnetic moments of impurity atoms is realized by means of RKKY
interaction which is replaced by the effective magnetic field,
whereupon system properties are described with the help of the
distribution function of local values of the field arising as a
result of magnetic ions' coupling with their own surroundings. In
real systems, the scattering of those fields proves to be so
substantial that RKKY interaction makes the magnetic ordering
possible at lower temperatures only (as compared to those predicted
by the traditional mean field theory). Dependencies of magnetic
properties on the carrier concentration turn out to be unusual, as
well.

In~\cite{7,8,9}, the expression has been derived for the energy
$w(r)$ of indirect RKKY interaction for two parallel spins  ${\bf
S}_1$, ${\bf S}_2$ of magnetic ions spaced at the distance $r$ in
the \emph{two-dimensional} system with \emph{degenerated} carriers:
 \be\label{1}
w(r)=-\frac{m}{4\pi\hbar^2}\left(\frac{J_{\rm ex}}{N}\right)^2\!
F(r)\,{\bf S}_1{\bf S}_2,\quad F(r)=-\int\limits_0^{k_F}
kN_0(kr)J_0(kr)dk,
 \ee
where $J_{\rm ex}$ is the exchange energy for interaction of a spin
with a free charge carrier of the mass $m$, $N$ is the concentration
of lattice atoms ($N=1/a^{2}$ for the square lattice of the period
$a$); $J_{0,1}$ и $N_{0,1}$ are Bessel functions. To generalize that
result to the case of the arbitrary degeneracy (with the Fermi
energy $\varepsilon_F$ of any sign and value) it is sufficient to
introduce the Fermi distribution function in the integrand (\ref{1})
and extend the integration over the interval $0<k<\infty$~\cite{6}:
 \be\label{3}
F(r,T)=-\frac{1}{r^2}\int\limits_0^\infty\frac{yN_0(y)J_0(y)dy}
  {1+\exp[(\hbar^2y^2/2mr^2-\varepsilon_F)/k_BT]}.
 \ee

The behavior of the function (\ref{3}) is determined not only by the
temperature as such but also by the temperature dependence of the
Fermi energy. The latter circumstance has been ignored in~\cite{6}.
In the framework of the standard two-dimensional band and under the
invariable carrier concentration, the ratio
$\eta=\varepsilon_F/k_BT$ ($\eta<0$ refers to the non-degenerated
carriers) is defined by the relation
 \be\label{eta}
e^{\eta(T)}=e^{\pi\hbar^2n/mkT}-1
  \ee
that predicts negative $\eta$-values at $T=100$ K if
$n\lesssim10^{12}$ cm$^{-2}$. From here, e.g. for the values
$n=3\cdot10^{12}$ cm$^{-2}$, $m=0.5m_0$, $a=4$\AA\, corresponding to
the experiments~\cite{3,4}, the temperature dependence $\eta(T)$
results which is shown in the inset of Fig.~1. In fact, it is seen
that in the real experiment, carriers could not be considered as
degenerated (in~\cite{3,4} actual temperatures amount to 170 K).

Taking into account (\ref{eta}), the expression (\ref{1}) could be
written in the form
 \be\label{4}
w(\rho)=-J_{\rm eff}\, \phi(\rho,\tau)\,{\bf S}_1{\bf S}_2,
 \ee
where $J_{\rm eff}=(ma^2/4\pi\hbar^2 )J_{\rm ex}^2$,
 \be\label{5}
\phi(\rho,\tau)=-\frac{1}{\rho^2}\int\limits_0^\infty\frac{yN_0(y)J_0(y)dy}
  {1+\exp(y^2/\rho^2\tau-\eta)}=
-\frac{1}{\rho^2}\left[e^{2\pi^2\gamma x/\tau}-1
 \right]\int\limits_0^\infty\frac{yN_0(y)J_0(y)dy}
  {e^{2\pi^2\gamma x/\tau}+e^{y^2/\rho^2\tau}-1},
  \ee
$\rho=r/a$ is the reduced separation between interacting impurities,
$\tau=2\pi ma^2k_BT/\hbar^2$ is the reduced temperature\footnote{For
GaMnAs $\tau\approx10^{-3} T $[K].}, $x=N_\mu/N$ is the relative
concentration of magnetic impurities. Under the strong degeneracy
($\eta\gg1$), Eq. (\ref{5}) reproduces the result~\cite{7}, while in
the case of non-degenerate carriers it is similar to the relevant
\emph{three-dimensional} result obtained in~\cite{10}. In the most
actual case of the intermediate degeneracy (see below), the function
(\ref{5}) has been estimated by numerical calculations.

In Fig. 1, dependencies $\phi(\rho)$ are shown calculated with
(\ref{5}) for different values of the carrier concentration
$x\gamma$ (the actual range of $x\gamma$-values is determined by
conditions of the concrete experiment). Particularly, it is clear
that at low enough carrier concentrations (corresponding to $\eta\le
0$) oscillations of the function $\phi(\rho)$ disappear and its sign
corresponds to the ferromagnetic interaction.

For simplicity we use the Ising model corresponding to $S=1/2$ and
leading, as is known, to qualitatively correct results at $S\gtrsim
1$, as well. Appropriate generalization does not meet some principal
difficulties.

Let the system consisting of randomly arranged and oriented Ising
spins be in the state characterized by the average reduced
magnetization $0\leqslant j\leqslant1$. The total interaction energy
$W=\sum_i w_i$ of a given spin ${\bf S}_1$ with other spins ${\bf
S}_i$ ($i=2,3,\ldots$) is a random value which we shall define by
the effective local magnetic field $H=-W/\mu$
($\mu=g\mu_B\sqrt{S(S+1)}$) and describe by the distribution
function $F(j;H)$ depending on the average concentration $N_\mu$ of
effective magnetic ions and the reduced system magnetization
$j=2\xi-1$ where $\xi$ is the average fraction of spins of
“magneto-active” ions directed “up”.

The randomness of the distribution of magneto-active impurities is
restricted  only by the necessity to place them in the fixed
locations (sites) of the matrix lattice. For  strongly diluted
systems, that restriction is not significant and the distribution
function could be found by the Markov's method~\cite{15}, according
to which
 \be\label{104}
F(j;H)\!=\!\frac{1}{2\pi}\! \int\limits_{-\infty}^\infty
\!\!A(q)\exp(-iqH)dq,\,\, A(q)\!=\!\lim _{N_{\rm
max}\to\infty}\!\left[\sum\limits_{\zeta=\pm1}\,
\int\limits_{0}^{r_{\rm max}}\!\!e^{iqh_\zeta(r,\,\zeta)}
\kappa_\zeta(\zeta)\kappa_r(r)dr\right]^{N_{\rm max}},
 \ee
where $h_\zeta(r,\zeta)=-\zeta w(r)/\mu$  is the field generated at
the origin by the spin spaced at the random distance $r$ from it.
The random parameter $\zeta$ takes values $\pm1$ (with probabilities
$\xi$ and $(1-\xi)$, accordingly) and determines the direction of
the remote spin, $\kappa_\zeta(\zeta)$ and $\kappa_r(r)$ are
distribution functions for random values of the parameter $\zeta$
and the distance $r$, $N_{\rm max}=\pi r_{\rm max}^2 N_\mu$ is the
number of magneto-active impurities in the circle of the radius
$r_{\rm max}$ over whose area the integration is performed.

In the spirit of the mean field theory, $\zeta$-distribution could
be written as
 \be\label{zeta}
\kappa_\zeta(\zeta)=[(1-\xi)\delta(\zeta+1)+\xi\,\delta(\zeta-1)].
 \ee
As for the $r$-distribution, $\kappa_r(r)$, one has, in principle,
to account for the clustering effect originating from the mutual
attraction experienced by near Mn-atoms in GaAs-lattice and
resulting in their non-random (correlated) spatial distribution.
Uniform non-correlated $r$-distribution would be
 \be\label{105}
\kappa_r(r) =
 \left\{
 \begin{tabular}{ll}
 $(2r/(r_{\rm max}^2-r_{\rm min}^2)$,&
$r>r_{\rm min}$\\
0&$r<r_{\rm min},$\\
 \end{tabular}
\right.
 \ee
where   the existence of the minimal separation $r_{\rm min}$
between magnetic ions  (conditioned by the embedding of impurity
atoms in the lattice of semiconductor matrix) is accounted
for\footnote{The minimal possible distance between magnetic-active
Mn-ions substituting for Ga-atoms in the zinc-blende AsGa-lattice
equals
 $a=a_0/\sqrt{2}\approx 4$\,\AA\, where
$a_0=5.7$\,\AA\, is the side of the cubic cell.}. To take into
account the correlation of Mn-atoms one has to add the correlation
function $g(r)$ in the right-hand side of that relation. This
function could be, in principle, calculated if the spatial
dependence of Mn-atoms interaction energy would be known. However,
the exact reason for the tendency of Mn-atoms to clustering is
unclear \cite{16}. At the same time,  Monte-Carlo calculations
\cite{17} show that impurity correlations have only small effects on
the ferromagnetic transition temperature of Ga$_{1-x}$Mn$_x$As
3D-system. Thus, though the consideration of the clustering is, in
principle, straightforward -- it is sufficient to introduce in the
integrand (\ref{105}) (and (\ref{106}), (\ref{108}), see below) the
correlation function $g(r)$ -- later on, we will use the simple
relation (\ref{105}) that corresponds to $g(r)=1$.

Substituting (\ref{zeta}), (\ref{105}) in (\ref{104}) one finds
 \be\label{106}
 A(q)=\exp[-2\pi  N_\mu C(q)],\quad
C(q)=\int\limits_{r_{\rm min}}^\infty\left\{1-\cos[qh(r)]-i\cdot
j\sin[qh(r)]\right\}r\, dr.
 \ee

Relationships (\ref{106}) do not lead to a simple analytical
expression for the distribution function $F_x(j;H)$. So, to
determine the latter we have used the “low $q$ approximation”, based
on the fact that in the inverse Fourier transform (\ref{104}) the
region of high $q$-values is not  important. In that approximation
 \be\label{107}
C(q)=P q^2-ij Q q,
 \ee
where
  \ba\label{108}\nonumber
 P=\frac{1}{2}\int\limits_{r_{\rm min}}^\infty\!\! h^2(r)r dr=
 (J_{\rm eff}/\mu)^2a^2\,\phi_P(\rho_{\rm min}),\quad
\phi_P(\tau)=\frac{1}{2}\int\limits^\infty_{\rho_{\rm min}}\!\!\phi^2(\rho)\rho d\rho,\nonumber\\
 \\
Q=\int\limits_{r_{\rm min}}^\infty\!\! h(r)rdr=
 (J_{\rm eff}/\mu)a^2\,\phi_Q(\rho_{\rm min}),\quad
\phi_Q(\tau)=\int\limits^\infty_{\rho_{\rm min}}\!\!\phi(\rho)\rho
d\rho,\nonumber
 \ea
$\rho_{\rm min}\equiv r_{\rm min}/a=1$.

Substituting~(\ref{107}), (\ref{106}) in (\ref{104}) we find that in
the considered approach the distribution $F(j;H)$ is described by
the shifted (relative to $H=0$) Gauss function\footnote{The lattice
model~\cite{13} used in~\cite{6} corresponds to $\delta$-like
distribution function $F(j,H)=\delta(H-jH_j)$ where $\mu H_j=\sum_i
N_i w(r_i)$; $N_i$, $r_i$ are the number of $i$th nearest neighbors
and their distance.}
 \be\label{109}
F(j;H)=
 \frac{\ds 1}{\ds \sqrt{2\pi}\sigma} \exp\left[-\frac{\ds (H-jH_j)^2}{\ds
 2\sigma^2}\right],
 \ee
  \be\label{110}
 H_j=2\pi  N_\mu Q\propto  N_\mu,
 \quad \sigma=(2\pi  N_\mu P)^{1/2}\propto  N_\mu^{1/2}.
 \ee
\indent The position of the maximum ($H=jH_j$) of the distribution
is determined by the parameter $Q$ and depends linearly on the
system magnetization $j$ while the distribution width $\sigma$ is
defined by the parameter $P$ and does not depend on $j$.  The
positive sign of $H_j$ means that the average direction of the
effective magnetic field coincides with the direction of the average
magnetization, that is the field is, on average, promotes the
\emph{ferromagnetic} ordering of magnetic moments.

Temperature dependencies $H_j(\tau)$ of the exchange field defined
by the relation (\ref{110}) are shown in Fig. 2. As illustrated in
the inset of that Figure,  $H_j$-value  peaks at $x\gamma\sim0.01$.
It is just this condition which results in the maximum Curie
temperature (see below).

Relations (\ref{110}) for the shift $H_j$ of the distribution
function $F(j; H)$ and its broadening $\sigma$ could be rewritten in
the form
 \be\label{111}
 H_j=(J_{\rm eff}/\mu)\left[2\pi x\, \phi_Q(\tau)\right],
\quad\sigma=(J_{\rm eff}/\mu)\left[2\pi x
\,\phi_P(\tau)\right]^{1/2}.
 \ee
It follows herefrom
 \be\label{112}
 H_j/\sigma=\left(2\pi x\right)^{1/2}\psi(\tau),
  \quad \psi(\tau)=
    \frac{\phi_Q(\tau)}{\,\,\,[\phi_P(\tau)]^{1/2}\,}.
 \ee

As is shown below, the ferromagnetic ordering is possible under the
condition $H_j/\sigma>\sqrt{\pi/2}$ or
 \be\label{113}
x^{1/2}\psi(\tau)>1/2.
 \ee
Thus, in addition to the “material” parameters $x=N_\mu/N$ and
$\gamma=n/N_\mu$,  the function $\psi(\tau)$ defining the ratio
$H_j/\sigma$ of Gauss function parameters becomes to be crucial in
the considered problem. As an example, temperature dependencies of
that ratio for $x=0.1$ and various $\gamma$-values are shown in
Fig.~3. In the inset, the dependencies of the maximum Curie
temperature $\tau_{\rm C}^{\rm max}$ on $\gamma$-value following
from (\ref{113}) is displayed (for various $x$-values).

In the traditional mean field theory, the distribution function is
$\delta$-like one for any magnetization $j$:
$F(j;H)=\delta[H-jH_j]$. It is evident, that the broadening of that
distribution in a random system prevents ferromagnetic ordering. The
magnetization of such a disordered system has to be calculated
taking into account the scattering of local interaction energies $H$
by means of the straight-forward generalization of the equation
$j=\tanh[\mu H(j)/kT]$ referring to the regular Ising system:
 \be\label{114}
j=\int\limits_{-\infty}^\infty \tanh\left[\frac{\mu H}{kT}\right]
F(j;H)dH.
 \ee

Using the expression (\ref{109}) for the distribution function
$F(j;H)$ one gets the equation generalizing the standard mean field
one:
 \be\label{115}
j=-\frac{1}{\sqrt{2\pi}}\left(\frac{H_j}{\sigma}\right)\int\limits_{-\infty}^\infty
\tanh\left(\frac{u}
{\theta}\right)\exp\left[-\frac{1}{2}\left(\frac{H_j}{\sigma}\right)^2(u-j)^2\right]\,du,
 \ee
where  $\theta=kT/\mu H_j=\tau/xI^2\phi_Q(\tau)$,
$I=\sqrt{\pi}J_{ex}(\hbar^2/ma^2)^{-1}$ is the reduced strength of
the interaction\footnote{Presently accepted value $J_{pd}=0.15$
eV$\cdot$nm$^3$~\cite{2} (for GaMnAs) results in $I\approx1$.}. That
equation predicts the phase diagram of the system, temperature
dependencies of its magnetization (in ferromagnetic phase) and
susceptibility (in paramagnetic phase), as well as the dependence of
Curie tempera\-ture $\theta_{\rm C}$ on the interaction strength,
the relative magnetic ion concentration $x=N_{\mu}/N$ and the
relative free carrier concentration $\gamma=n/N_{\mu}$.

To clarify under what conditions that equation has the solution
corresponding to the ferromagnetic state ($j>0$) notice that in the
vicinity of  Curie temperature where the magnetization is small
($j\to 0$), it follows from (\ref{115})
 \be\label{116}
\sqrt{\frac{2}{\pi}}\left(\frac{H_j}{\sigma}\right)^3\int\limits_{0}^\infty
\tanh\left(\frac{u}{\theta}\right)\exp\left[-\frac{1}{2}\left(\frac{H_j}{\sigma}\right)^2
u^2\right]u\,du=1.
 \ee
 The integral in (\ref{116})  peaks at
$\theta=\tau=0$ and its maximum value equals $(\sigma/H_j)^2$. It
follows herefrom that the ordered state is only possible under the
condition
 \be\label{117}
\frac{H_j}{\sigma}> \sqrt{\frac{\pi}{2}},
 \ee
which means that the effective RKKY-field proportional to $H_j$ has
to “overpower” not only the thermal disordering but also the
scattering of local fields proportional to $\sigma$. The upper
boundary  $\tau_{\rm C}^{\rm max}$ of the temperature range where
the cited condition is satisfied determines the maximum attainable
temperature of the ferromagnetic ordering at \emph{infinite}
interaction energy  ($I\to\infty$). It could be derived from the
condition $2x^{1/2}\psi(\tau)=1$. The existence of the maximum Curie
temperature is associated with the lifting of the carrier
degeneration at high temperatures and, as a consequence, the
finiteness of the effective energy of inter-impurity interaction
even at  $I\to\infty$.

Curie temperature at the finite interaction energy could be
determined  by solving the equation (\ref{116}). Relevant
non-monotone dependencies $\tau_{\rm C}(\gamma)$ are displayed in
Fig.~4. It is clear that the relative carrier concentration
$\gamma=n/N_\mu$ ambiguously influences Curie temperature in
accordance with the non-monotone dependence of the exchange field
$H_j$ on $\gamma$ (cf. Fig. 2). To compare, the dashed line in Fig.
4 reproduces the dependence $\tau_{\rm C}(n)$ presented in~\cite{6}
and obtained in the framework of the standard mean-field theory.

The optimal carrier concentration occurs to be on the order of
$10^{12}$ cm$^{-2}$, that is significantly lower than the value
$n\sim10^{14}$ cm$^{-2}$ predicted in~\cite{6}. In addition, Fig. 5
demonstrates there is a threshold value of the interaction strength
$I$ to drive the system in the ferromagnetic state. This is to be
contrasted with the result of the standard mean-field theory which
predicts no such a threshold.

In conclusion,  conditions of establishing the ferromagnetic state
and its parameters in quasi-two dimensional semiconductor systems
with magnetic impurities coupled via RKKY interaction have been
studied in the paper. As distinct from~\cite{6}, two new important
factors have been included in the consideration, allowing for the
spatial disarray of interacting magnetic impurities, and the
temperature dependence of the carrier degeneracy. It has been
demonstrated that both factors complicate transition of the system
into the ferromagnetic state: disorder of the impurities arrangement
reduces the Curie temperature (as compared to the regular system)
while lifting the degeneracy of carriers makes the Curie temperature
finite even in the extreme case of the infinitely strong
interaction. Besides, the concentration dependence of the transition
temperature occurs to be non-monotone and there is a threshold
interaction strength to drive the system in the ferromagnetic state.

This work has been supported by Grant No. 06-02-116313 of the
Russian Foundation of Basic Researches.

\renewcommand{\refname}{\centerline{\rm \small\bf References}\vspace{5mm}}

\newpage
\centerline{\bf Captions}
\bigskip
\bigskip

Fig. 1. Spatial dependencies $\phi(\rho)$ of the RKKY interaction
energy for magnetic impurities  in the two-dimensional system with
various carrier concentrations degeneracy determining by the
parameter $x\gamma$ (notice, different scales above and under the
Y-axis break). In the insert -- the temperature dependence
$\eta(T)$ corresponding to conditions of the experiments~\cite{3,4}.\\

Fig. 2. Temperature dependencies $H_j(\tau)$ of effective exchange
field for the two-dimensional system with various concentrations of
magnetic impurities  ($x=N_\mu/N$) and free carriers
($\gamma=n/N_\mu$). In the insert -- the maximum attainable $H_j$-value. \\

Fig. 3. Temperature dependencies of the ratio $H_j/\sigma$ of Gauss
distribution function parameters  for  two-dimensional system with
the concentration of magnetic impurities $x=0.05$ at various carrier
concentrations $\gamma$. In the insert -- maximum Curie temperature
$\tau_{\rm C}^{\rm max}$ that could be attained at
given $x$- and $\gamma$-values.\\

Fig. 4. Dependencies $\tau_{\rm C}(n)$ of Curie temperature  on the
carrier concentration for two-dimensional system with the
concentration of magnetic impurities $x=0.1$ for various interaction strengths $I$.
The dashed line is the result having presented in~\cite{6} and corresponding to $I\approx1$. \\

Fig. 5. Dependencies $\tau_{\rm C}(I)$ of Curie temperature   for
two-dimensional system with the concentration of magnetic impurities
$x=0.1$ at various carrier concentrations determined by the parameter $\gamma$. \\

\newpage


\begin{thebibliography}{30}
\bibitem{1} H. Ohno, Science, {\bf 281}, 951 (1998).

\bibitem{2} T. Dietl, H. Ohno, F. Matsukara, J. Cibert, D. Ferrand, Science, \textbf{287},
             1019 (2000).
\bibitem{3} A.M. Nazmul, S. Sugahara, M. Tanaka, Phys. Rev. B, {\bf
67}, 241308(R) (2003).

\bibitem{4} A.M. Nazmul, T. Amemiya, Y. Shuto,S. Sugahara, M. Tanaka, Phys. Rev. Lett., {\bf
95}, 017201 (2005).

\bibitem{5} R. Morya, H. Munekata, J. Appl. Phys., {\bf 93}, 4603
(2003).

\bibitem{6} D. Priour, Jr., E.H. Hwang, S. Das Sarma, Phys. Rev. Lett., {\bf 95},
037201 (2005).

\bibitem{7} B. Fisher, M. Klein, Phys. Rev., {\bf 11}, 2025 (1975).

\bibitem{8} D. N. Aristov, Phys. Rev B, {\bf 55}, 8064 (1997).

\bibitem{9} V. Litvinov, V. Dugaev, Phys. Rev B, {\bf 58}, 3584 (1998).

\bibitem{10} B.V. Karpenko, A.A. Berdyshev, Sov. Phys. Solid State, {\bf 5},
2494 (1964).

\bibitem{11} A.A. Berdyshev, Sov. Phys. Soloid State, {\bf 8}, 1104
(1966).

\bibitem{12} M.I. Durby, J. Phys. D: Appl. Phys., {\bf 3}, 1491
(1970).

\bibitem{13} D.J. Priour, Jr., E.H. Hwang, S. das Sarma, Phys. Rev. Lett., {\bf 92}, 117201 (2004).

\bibitem{14}  Meilikhov E.Z., Farzetdinova R.M., JMMM, {\bf
293}, 793 (2005).

\bibitem{15}  S. Chandrasekhar, Rev. Mod. Phys, {\bf 15}, 1 (1943).

\bibitem{16}  P. Mahadevan, J.M. Osorio-Guillen, and A. Zunger, Appl. Phys. Lett., {\bf 86},
172504 (2005).

\bibitem{17}  D.J. Priour,  S. das Sarma, arXiv:cond-mat/0509614 v2
22Feb2006.
\end{thebibliography}
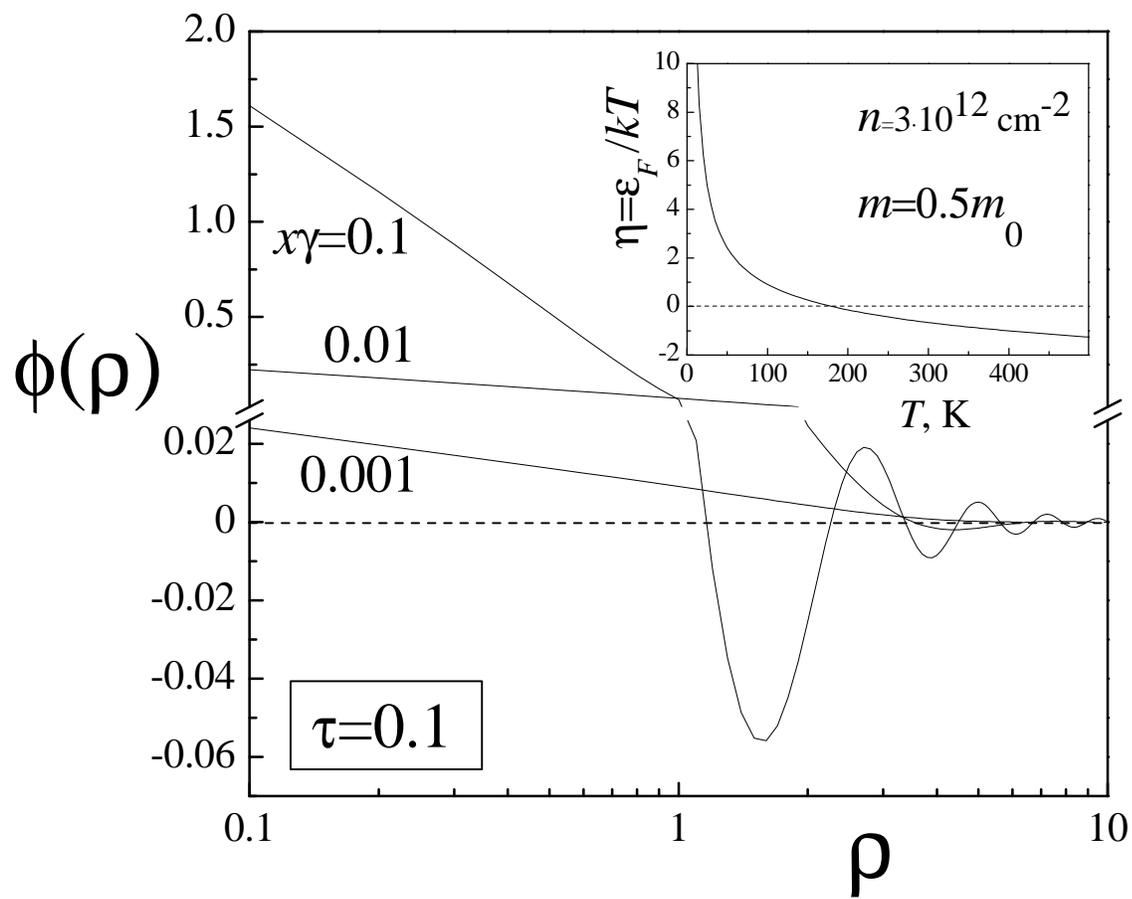
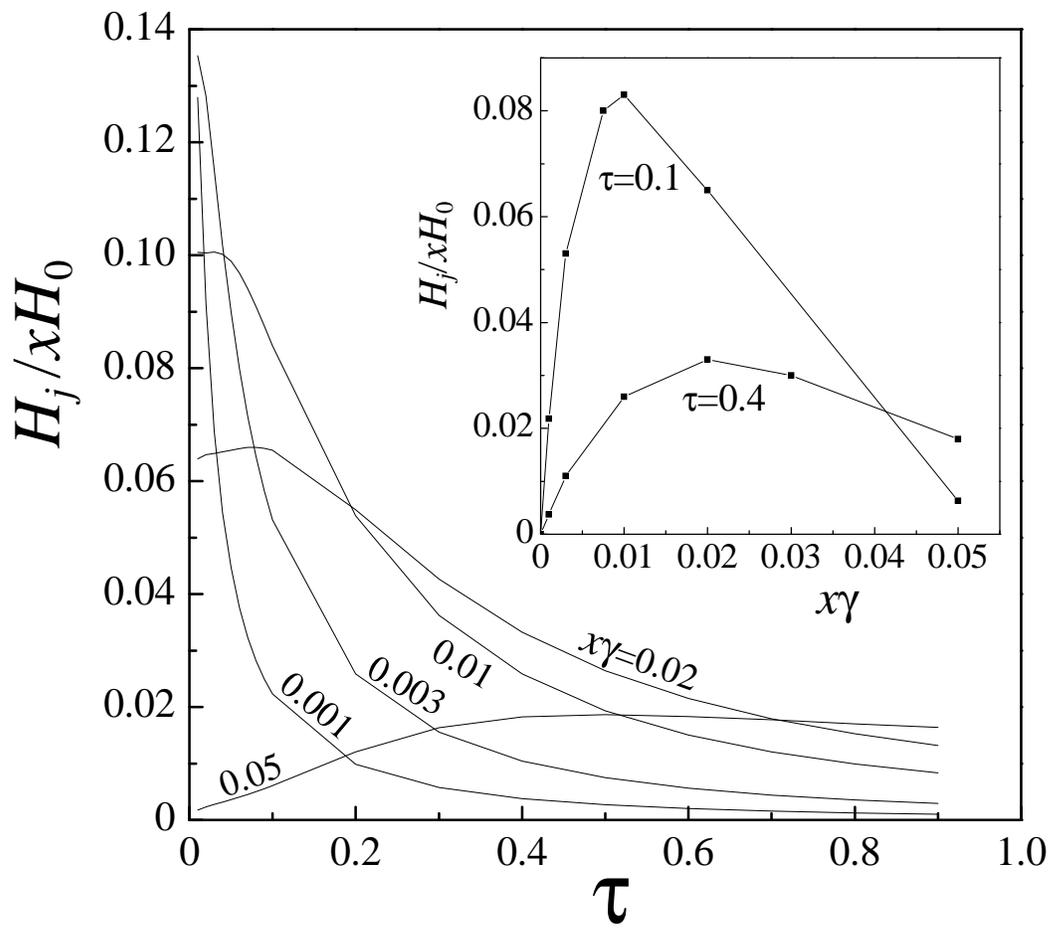
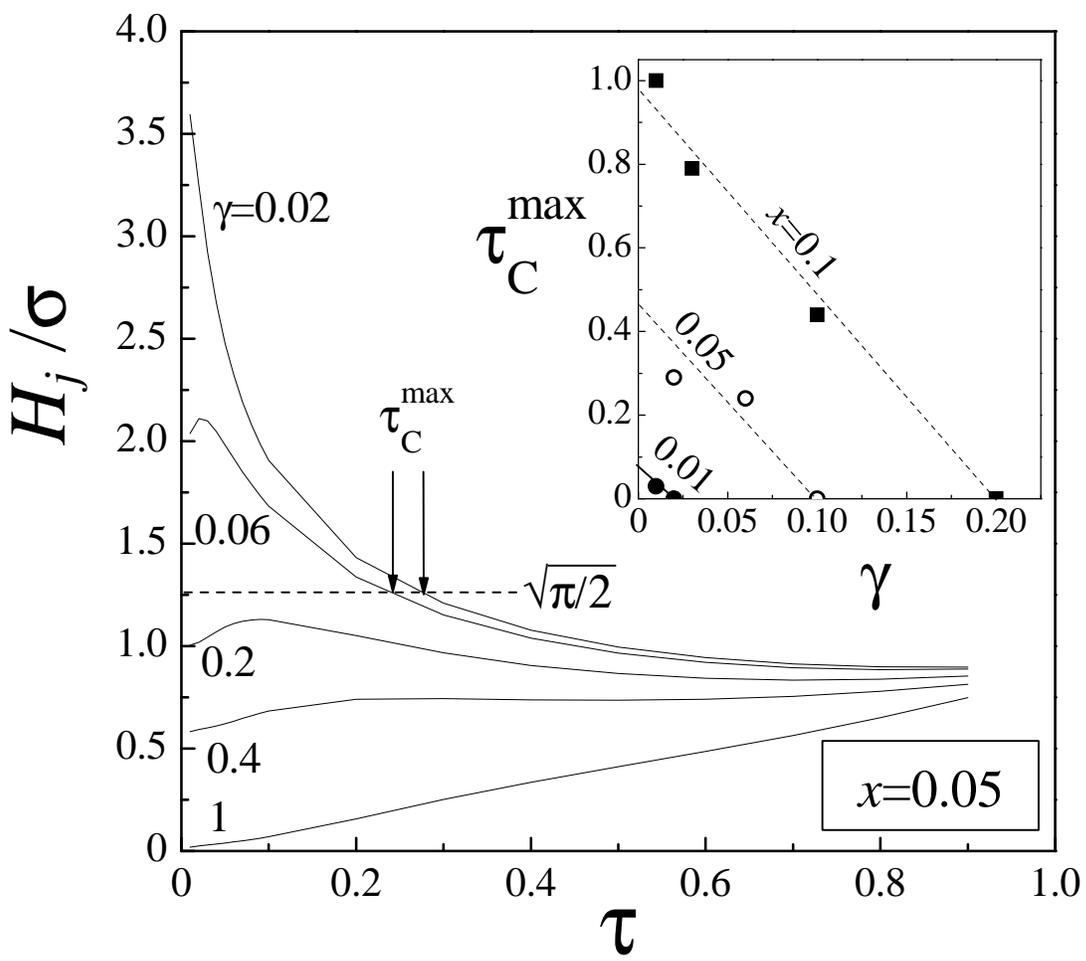
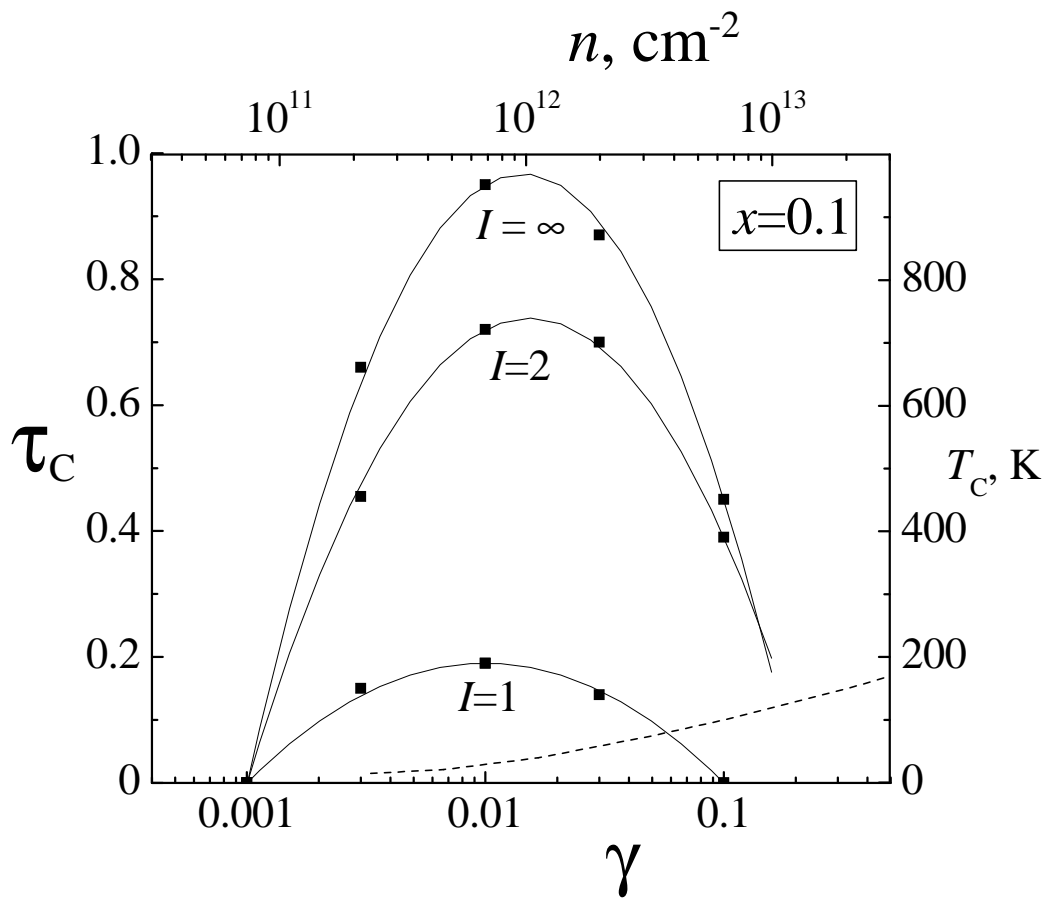
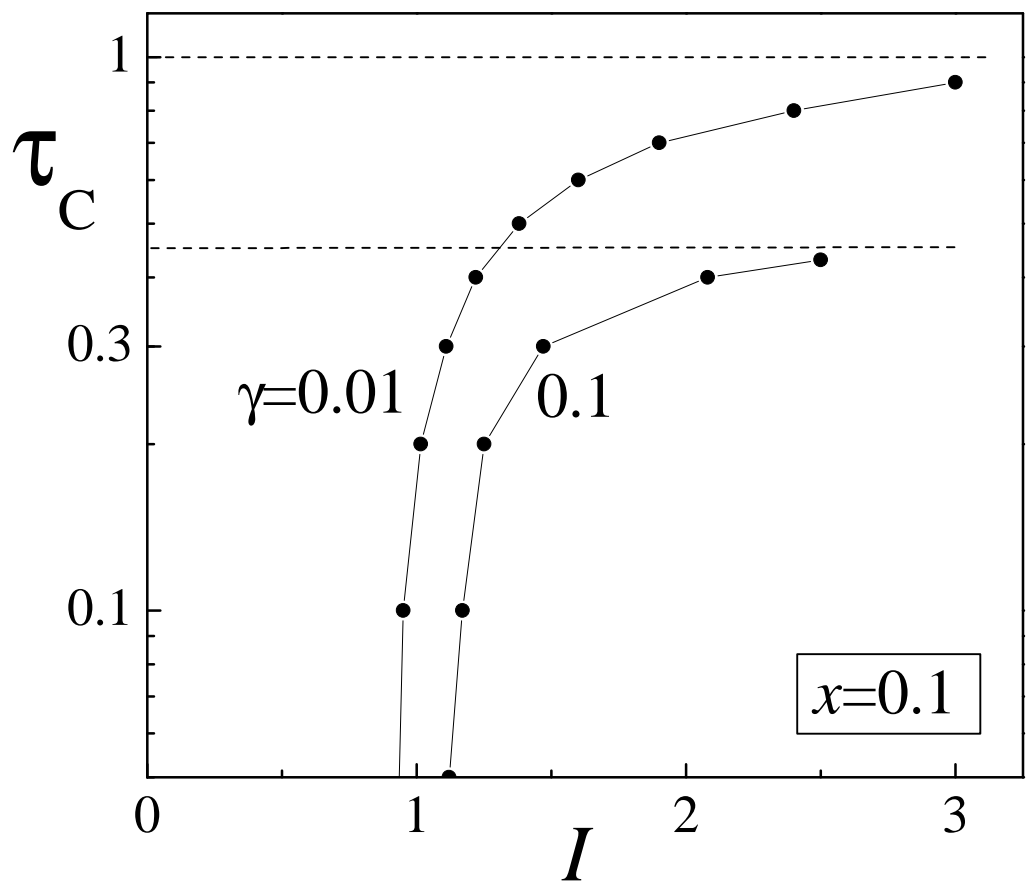
\end{document}